\newcommand{\Pl}{\partial}
\newcommand{\ts}{\textstyle}

\newcommand{\fpar}[2]{\frac{{\ts \Pl \/ #1}}{{\ts \Pl \/ #2}}}
\newcommand{\nder}[3]{\frac{{\ts d^{#1} \/ #2}}{{\ts d \/ #3^{#1}}}}

\newcommand{\bee}{\begin{equation}}
\newcommand{\ene}{\end{equation}}
\newcommand{\beea}{\begin{eqnarray}}
\newcommand{\enea}{\end{eqnarray}}

\documentclass[prb,aps,twocolumn]{revtex4}
\usepackage{dcolumn}
\usepackage{graphicx}
\usepackage{latexsym}
\usepackage{amsmath}
\usepackage{amssymb}

\begin{document}

\title{Phase Mixing of Large Amplitude Relativistic Electron Plasma Oscillation With Inhomogeneious Ion Background }
\author{Mithun Karmakar,$^{1,2}$ Chandan Maity,$^3$, and  Nikhil Chakrabarti $^{1,2}$}
\affiliation{$^1$ Saha Institute of Nuclear Physics, 1/AF Bidhannagar, Kolkata-700064, India
\\$^2$ Homi Bhaba National Institute, Training School Complex, Anushakti Nagar, Mumbai 400085, India
\\$^3$ Vivekananda Mahavidyalaya, Haripal, Hooghly-712405, India.}
\begin{abstract}
{Phase mixing of relativistic large amplitude nonlinear plasma wave in presence of a time independent space periodic ion density
profile has been investigated. Inhomogeneous ion along with the relativistic variation of electron mass make
the characteristic frequency of the wave to acquire a space dependency and 
thereby it breaks at arbitrarily small amplitude due to phase mixing. An approximate space time dependent 
solution is obtained in the weakly relativistic limit by Bogoliuboff and Kryloff method of 
averaging. We find that the change in the  
ion density perturbation and also the relativistic electron mass variation have significant effect in modifying the time at which
phase mixing occurs.}
\end{abstract}

\maketitle

 \begin{section}{Introduction}
  
  
  The onset and evolution of relativistic electron plasma wave has been a topic of
  fundamental interest to the plasma physicists during the last few decades.\cite{akhiezer,davidson1,davidson2,davidsonbook,stenflo,brodin,xu,zhou} 
  An increasing attention has been given to unveil the detailed understanding of the 
  nonlinear plasma wave phenomena due to its relevance in the broad range
  of physical processes related to astrophysical plasma, plasma based accelerators, plasma heating etc.\cite{tajima,joshi,modena}

  The properties of nonlinear relativistic electron plasma wave have been investigated first by
  Akheizer and Polovin. \cite{akhiezer}
  They have reported an analytical estimation of the limiting value of the electric field that can be supported by 
  the plasma for the longitudinal plasma wave. This limiting value now well recognized as 
  the wave breaking amplitude is one of the key features in plasma heating process as well as in the determination of 
  the maximum energy gain in the plasma based acceleration process. Dawson have first introduced the notion of the 
  non-relativistic wave breaking to describe the limiting amplitude of the plasma wave and demonstrated the breaking phenomena
  in terms of sheet model. \cite{dawson} Subsequently, an exact space time dependent
  Lagrangian solution for non-relativistic nonlinear plasma wave has been obtained by Davidson and Schram.\cite{davidson1}
  The solution presented by them shows that electron density does not show explosive behavior
  unless the perturbation amplitude exceeds half of the equilibrium density. Thereafter, thermal corrections are 
  incorporated in the expression of the breaking electric field amplitude by Coffey \cite{coffey} and also by Katsouleas and Mori.\cite{katsouleas}

%
%
%

     Most of these studies on wave breaking of nonlinear electron plasma wave have been done with the simplifying assumption of 
     a static homogeneous ion background. However, if the ion density becomes inhomogeneous, 
     the characteristic frequency of the plasma wave acquire space dependency causing 
     different elements of the plasma to oscillate with different frequencies. This leads to 
     loss of wave coherency resulting breaking of the wave via phase mixing.\cite{sengupta2,sengupta3,pramanik,pramanik2}
     Phase mixing can also be possible even without the density inhomogeneity which occurs due to relativistic electron mass variation.
     This nonlinear effect has been shown to play a crucial role 
     for the occurrence of relativistic bursts discussed in the theoretical analysis of Infeld and Rowland. 
     They have shown that except for a particular choice of initial condition,
     an explosive behavior of the excited wave is always observed and the wave breaks at arbitrarily small amplitude 
     long before reaching Akhiezer Polovin limit.\cite{infeld1}
%
%

      There exists several works where the inhomogeneity in the ion density has been considered.\cite{nappi}
      In 1989, Infeld et. al. have investigated the phase mixing process in presence of 
      sinusoidal time stationary ion density inhomogeneity in the non-relativistic plasma system.\cite{infeld2}
      Later, similar analysis has been done for the nonlinear Langmuir oscillation against single-ion pulse or cavity background.\cite{infeld3}
      Such theories are aimed to unravel the basic understanding of non-relativistic wave dynamics in inhomogeneous plasma, 
      but of course give no indication of the relativistic effects in presence of background ion inhomogeneity.
      Admittedly, the development of the fully relativistic theory of nonlinear electron plasma wave
      in presence of ion density inhomogeneity encounters with significant mathematical difficulties. 
      So any method which is capable of considering both the 
      relativistic effect and inhomogeneity in the ion concentration can be of much importance. 
      Such an attempt has been made here in this paper, where by using an acceptable simplifying assumption, we have 
      obtained an approximate space time dependent solution in the weakly relativistic limit.
      Here, the basic difference from most of the earlier analyses is that in presence of 
      ion inhomogeneity, instead of a uniform initial electron density profile we have 
      considered a finite amplitude perturbation given to it. Such type of initial condition has been adopted earlier
      by Kaw et al. in order to investigate coupling effect of electron 
      plasma oscillation with other waves by ion density fluctuation.\cite{kaw}
      They have assumed a time independent ion fluctuations along with a finite amplitude sinusoidal perturbation 
      given to electron density as initial condition. 

      In section (II) the basic equations describing the phase mixing process of relativistic nonlinear
      plasma wave are given and to solve these equations Lagrangian variables are introduced.
      Section (III) describes the the approximate solution obtained by the Bogoliuboff and Kryloff method of 
      averaging.\cite{bogoliubov} Here we have also recovered some well known results in different limiting cases.
      Finally in the conclusion section we have summarized our results.

 \end{section}

 \begin{section}{The basic equations and its solution}

   The nonlinear relativistic dynamics of the large amplitude electron plasma wave can be described by the
   following one dimensional Eulerian fluid equations:

   \beea
\partial_t{n_e}+ \partial_{x} (n_e v_e)=0,
\label{eq1}
\enea
\beea
\partial_t{p_e}+ v_e\partial_x {p_e}=-{e{E}}/{m_e},
\label{eq2}
\enea
\beea
\partial_x{E}=4 \pi e (n_i-n_e),
\label{eq3}
\enea

  The symbols used to describe different dynamical variables have their usual meanings. We assume that, in the equilibrium plasma state, 
  a pre-existing ion wave induces an inhomogeneity in the background ion density.
  For the sake of simplicity, we consider a time independent but space periodic ion density profile of the form, 
\beea
n_i(x,t)=n_0(1+\delta_i \cos k_i x),
\label{eq4}
\enea
where $\delta_i$ and $k_i$, respectively, denote the amplitude and wavenumber of the ion inhomogeneity. Moreover, we take the perturbation in the electron density as
\beea
n_e(x,0)=n_0(1+\delta_e \cos k_e x),
\label{eq5}
\enea
where $\delta_e$ and $k_e$, respectively, signify the strength and inverse of scale length of the perturbation. Thus, an initial charge imbalance between electron and ion creates an electric field which can be obtained from Eq. (\ref{eq3}) as
\beea
E(x,0)=4 \pi e n_0\left({\delta_i}{k_i}^{-1} \sin k_i x-{\delta_e}{k_e}^{-1} \sin k_e x\right).
\label{eq6}
\enea
This initial electric field drives the plasma system in a nonlinear electron plasma mode.

%
%
%
%
%
%
%
%
%
%
%
%
%
%
%

Next, to find an exact space-time dependent solution of the problem, we introduce Lagrangian coordinates ($\xi,\tau$) through an auxiliary variable $\psi$:
$\xi=x-\psi(\xi,\tau),\; \psi=\int_0^\tau v_e(\xi,\tau) d\tau,\; \tau=t.$
The basic equations in the transformed co-ordinate system simplify to the following forms,

\beea
\fpar{}{\tau}\left[n_e\left(1+\fpar{\psi}{\xi}\right)\right]=0 ;  \fpar{\psi}{\tau}=v_e,
\label{a9}
\enea

\beea
\fpar{p_e}{\tau}=-e E/m_e,
\label{a10}
\enea

\beea
\fpar{E}{\tau}=4 \pi e n_i v_e.
\label{a11}
\enea

The continuity equation simplifies to, 

\beea
n_e(\xi,\tau)=\frac{n_e(\xi,0)}{\left(1+\fpar{\psi}{\xi}\right)}
\label{a12}
\enea

The momentum equation and the electric field evolution equation with the prescribed initial conditions combined to give us,

\beea
\frac{\ddot{\psi}}{\left(1-\frac{\dot{\psi}^2}{c^2}\right)^{3/2}}+\omega_p^2\left[\psi + \frac{\delta_i}{k_i} \sin k_i (\xi + \psi)-\frac{\delta_e}{k_e} \sin k_e \xi \right]=0\nonumber\\
\label{a13}
\enea

%
%
%

 Now here we make an assumption that the perturbation length scale for the ion density is much higher
 than that of the electron length scale i.e. $k_i << k_e$. The assumption of large scale space variation of ion density 
 is quite justified due to the fact that ions are less mobile than the electron. In order to gain some insight into the relativistic 
 effects on the longitudinal plasma wave with a inhomogeneous ion background, it is sufficient and analytically
 tractable to retain terms up to first order in $k_i \psi$ in the following expansion,

\beea
\sin k_i(\xi +\psi) \approx \sin k_i \xi +\kappa (k_e \psi)\cos k_i \xi  \nonumber
\label{a14}
\enea
   
 where $\kappa=(k_i/k_e)<< 1$.

 Therefore, Eq.(\ref{a13}) can now be written as,

\beea
\ddot{\psi}+\omega_p^2\left[\psi +\frac{\delta_i}{k_i}\left\{\sin k_i \xi +
\kappa (k_e \psi)\cos k_i \xi \right\}\right] \left(1-\frac{3}{2}\frac{\dot{\psi}^2}{c^2}\right)\nonumber\\
-\frac{\omega_p^2\delta_e}{k_e} \sin(k_e \xi)\left(1-\frac{3}{2}\frac{\dot{\psi}^2}{c^2}\right) \approx 0  \nonumber
\label{a15}
\enea
   
  Introducing normalized variables $k_e \psi=\phi$, $k_e \xi=\bar{x}$, $\omega_p \tau =\bar\tau$ , 
this equation simplifies to

\beea
\nder{2}{\phi}{\bar\tau}+f_1 \phi -f_2 \phi \dot{\phi}^2-f_3 \dot{\phi}^2+f_4\approx 0,
\label{a16}
\enea
where $f_1=1+\delta_i \cos(\kappa \bar{x}),\; f_2=\frac{3}{2}\beta^2 [1+\delta_i \cos(\kappa\bar{x})],\; 
f_3=\frac{3}{2}\beta^2[(\delta_i/ \kappa) \sin \kappa \bar{x}-\delta_e \sin \bar{x}],\; 
f_4=(\delta_i/ \kappa) \sin \kappa \bar{x}-\delta_e \sin \bar{x} $ with $\beta=\omega_p/k_e c$.

%
%

  \end{section}

\begin{section}{Nonlinear solution : Bogoliuboff and Kryloff method}

In order to obtain a solution for the the second order nonlinear differential equation (Eq. {\ref{a16}}), 
we introduce a new variable $\chi=\phi+\frac{f_4}{f_1}$. With that transformation,
this equation takes the form as

\beea
\nder{2}{\chi}{\bar\tau}+f_1 \chi -f_3 \chi \dot{\chi}^2=0,
\label{eq15}
\enea

We now wish to find the solution for the above equation for small $\delta_i$. In that limit 
we can consider $f_3$ as a small parameter and thereby we obtain the solution of the 
above equation by the method of Bogoliuboff and Kryloff method of 
averaging\cite{bogoliubov} as,
       
\beea
\chi(\bar{x},\bar\tau)=\chi_0(\bar{x})\sin[\bar{\omega}\bar\tau+\theta(\bar{x})]
\label{eq15}
\enea
where\\
\beea
\bar{\omega}=(1+\delta_i \cos\kappa \bar{x})^{1/2}\left[1-\frac{3 \beta^2}{16}\frac{[(\delta_i/ \kappa) \sin \kappa \bar{x}-
\delta_e \sin \bar{x}]^2}{1+\delta_i \cos\kappa \bar{x}}\right].\nonumber
\label{eq15}
\enea
 
 It is clearly seen from this expression that the characteristic 
frequency of the nonlinear plasma wave acquire space dependency. As a result, different fluid elements
oscillate with different frequency and become out of phase with each other. The combined 
effect of inhomogeneous ion background along with the relativistic electron mass variation is 
responsible for this loss of wave coherency which leads to destruction of the plasma wave (wave breaking).
 
The full solution of the problem depends upon the two unknown functions $\chi_0(\bar{x})$ and $\theta(\bar{x})$ which can be determined 
by our prescribed initial condition viz. $\phi(\bar{x},0)=\dot{\chi}(\bar{x},0)=0$. This gives us $\theta(\bar{x})=\pi/2$,  and 
$\chi_0(\bar{x})=f_4/f_1$.
So finally we can write the approximate solution truncated to the second order in $\delta_i$ as:\\

\beea
\phi(\bar{x},\bar\tau)=f_4(1-\delta_i \cos\kappa \bar{x})[\cos(\bar{\omega}\bar\tau)-1].
\label{eq15}
\enea

%
%

   \begin{figure}[ht]
{\centering
{\includegraphics[width=2.5in,height=1.5in]{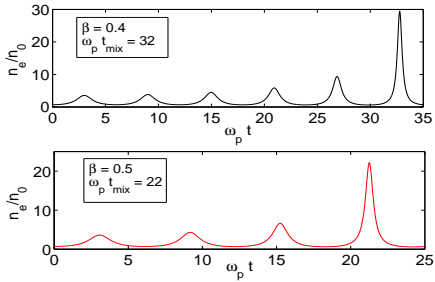}\par}}
\caption{Density profiles. Here $\delta_e=0.3$, $\delta_i=0.2$, $\kappa=0.1$ and $\bar{x} = \pi$.}
\label{fig:1}
\end{figure}

From the solution we calculate  $\fpar{\phi}{\bar{x}}$ and obtain 
the expression for electron density as,

\beea
n_e(\bar{x},\bar\tau)=\frac{n_0[1+\delta_e \cos \kappa\bar{x} ]}{1+A(\bar{x})(\cos \bar{\omega}\bar\tau-1)-B(\bar{x})\bar\tau\sin\bar{\omega}\bar\tau},\nonumber  
\label{eq15}
\enea 
where,         
\beea
A(\bar{x})=(\delta_i \cos\kappa \bar{x}-\delta_e \cos\bar{x})(1-\delta_i\cos\kappa \bar{x})+\delta_i \kappa f_4 \sin\kappa \bar{x};\nonumber
\label{eq15}
\enea
\beea
B(\bar{x})=-\frac{\delta_i \kappa f_4\sin\kappa \bar{x}}{2}(1-\delta_i\cos\kappa \bar{x})+\frac{\delta_i^2\kappa f_4}{8} \sin(2\kappa \bar{x})\nonumber\\
-\frac{3}{8}\beta^2f_4^2(\delta_i\cos\kappa \bar{x}-\delta_e \cos\bar{x}).\nonumber
\label{eq15}
\enea

\begin{figure}[ht]
{\centering
{\includegraphics[width=2.5in,height=1.5in]{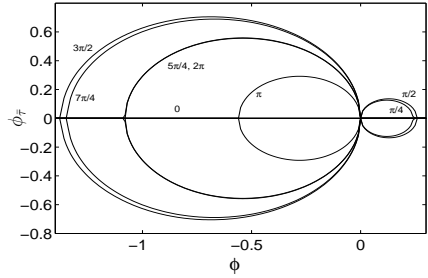}\par}}
\caption{Phase-space curves for different values of $\kappa \bar{x}$. Here $\beta=0.4$, $\delta_e=0.3$, $\delta_i=0.1$ and $\kappa=0.1$.}
\label{fig:2}
\end{figure}

From the solution of $\phi$ we can obtain the phase space plot from the equation below,

\beea
\dot{\phi}^2+\bar{\omega}^2\phi^2+2 f_4(1-\delta_i \cos\kappa\bar{x})\bar{\omega}^2 \phi=0.
\label{eq15}
\enea

where $\bar{\omega}$ is approximated as 
         
\beea
\bar{\omega}=1+\frac{\delta_i}{2}\cos(\kappa\bar{x})-\frac{\delta_i^2}{8}\cos^2(\kappa\bar{x})-\frac{3}{8}\beta^2f_4^2.
\label{eq15}
\enea

 It is evident from the phase portrait Fig. (\ref{fig:2}) that individual fluid elements execute periodic motion
 for each value of $\bar{k} \bar{x}$ and the time period of oscillation depends on $\bar{x}$. 
 Physically, these phase portraits exhibit the periodic motion of individual fluid elements.
As different fluid elements are characterized by different values of $\bar{x}$, 
in the present context, the time periods of their oscillations depend on $\bar{x}$. 
This is in contrast to the case of homogeneous plasma system where the period of 
oscillation is independent of $\bar{x}$. Due to this space dependency, as time goes on, 
neighboring fluid elements start to cross their trajectories which indicates an onset of 
fine scale mixing of oscillations, leading to the breaking of excited oscillations at a finite time.

Now the phase-mixing time can be estimated considering the point $\partial\phi/\partial\bar{x}=-1$ at which singularity in the electron density is observed. The expression for the approximate phase-mixing time is obtained as
\beea
\omega_pt_{mix}\simeq \nonumber\\
(\delta_i-\kappa \delta_e)^{-1}\left[\frac{3}{8}\delta_i^2-\frac{\delta_i}{2}-\frac{3\beta^2}{16\kappa^2}(\delta_i^2-2\delta_i\delta_e+\kappa\delta_e^2)\right]^{-1}
\label{eq24}
\enea

   \begin{figure}[ht]
{\centering
{\includegraphics[width=2.5in,height=1.5in]{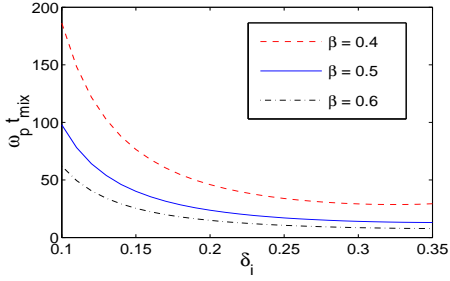}\par}}
\caption{Phase mixing time vs $\delta_i$ for different values of  $\beta$. Here $\delta_e=0.3$ and $\kappa=0.1$.}
\label{fig:3}
\end{figure}
  
Now we shall recover some earlier results in different limiting cases:

\subsection{Weakly relativistic homogeneous plasma ($\delta_i=0$)}
  
  If we consider phase mixing of electron plasma wave due to relativistic variation of electron mass in homogeneous plasma 
then we simply put $\delta_i=0$ in the density evolution expression to obtain:

\beea
n_e(\bar{x}\bar \tau)=\frac{n_0[1+\delta_e \cos \bar{k}\bar{x} ]}{1+\delta_e \cos \bar{x} [1-\cos (\bar{\omega_p}\bar \tau) -\frac{3}{8} (\delta_e \beta)^2 \bar \tau \sin^2 \bar{x} \sin(\bar{\omega_p \tau})]}\nonumber  
\label{eq15}
\enea
  
where,
         
\beea
\bar{\omega}_p = [1-(3/16)\beta^2 \delta_e^2\sin^2 \bar{x}].\nonumber
\label{eq15}
\enea

  This is the well known expression of the electron density obtained by Infeld for the homogeneous plasma 
  system in the weakly relativistic limit.\cite{infeld1} The occurrence of relativistic bursts as a result of relativistic mass variation
  has been discussed elaborately by him.
  In the limit of $\beta\rightarrow 0$ we get back the 
  density expression as obtained in the Davidson's Book. \cite{davidsonbook}

  The phase mixing time for the weakly relativistic wave takes the following form:\\
  
  \beea
\omega_pt_{mix}\simeq [(3/16)\beta^2\delta_e^3]^{-1}.
\label{eq24}
\enea

This is the same expression obtained by Sengupta et. al.\cite{sengupta2}
  
  \subsection{Nonlinear non-relativistic electron plasma oscillation in inhomogeneous ion background}
  
  In the non-relativistic limit we can recover the earlier results obtained to discuss phase 
  mixing in inhomogeneous plasma system. If we set $\beta=0$ we get back the results for the nonlinear electron plasma 
  wave when the perturbation on electron density is considered in the inhomogeneous ion background. The expression for density 
  take the form as (in the limit of small $\delta_i$),
  
\beea
n_e(\bar{x},\bar\tau)=n_0(1+\delta_e \cos \bar x)/D,
\label{eq21}
\enea
where,
\beea
D=1+f_4(\cos\bar\omega\bar\tau-1)+\delta_i {\kappa}f_3 \bar \tau \sin {\kappa}\bar{x} \sin\bar\omega\bar\tau,\nonumber
\label{e22}
\enea
where, $f_4=(\delta_i \cos {\kappa} \bar{x}-\delta_e \cos\bar{x})$.

   The phase mixing time is then,
   
   \beea
\omega_pt_{mix}\simeq \frac{2}{\delta_i(\delta_i-\kappa \delta_e)}
\label{eq24}
\enea

  When $\delta_e=0$, we recover the expression for the phase mixing time as obtained by Nappi et. al.\cite{nappi}
  as,

\beea
\omega_p t_{mix}\simeq \frac{2}{\delta_i^2}.
\label{eq24}
\enea

  Finally, we put $\delta_e=0$ and $\beta=0$ in the expression for 
  electron density. In the small $\delta_i$ limit we approximate $\phi$ by retaining terms upto second order in $\delta_i$
  as:

\beea
\phi(\bar{x},\tau)=[(\delta_i/\kappa)\sin \kappa\bar{x}-(\delta_i^2/2\kappa)\sin 2\kappa\bar{x}][\cos(\bar{\omega}\tau)-1]\nonumber\\
\label{eq15}
\enea

  Then, we find the expression for electron density as:

\beea
n_e(\bar{x},\bar \tau)=\frac{n_0}{R(\bar{x})[\cos(\bar{\omega}\bar\tau)-1]-S(\bar{x})\bar\tau \sin(\bar{\omega}\bar\tau)}
\label{eq15}
\enea

 where,          
\beea
R(\bar{x})=(\delta_i\cos \kappa\bar{x}-\delta_i^2\cos 2\kappa\bar{x}),
\label{eq15}
\enea
and 
  
\beea
S(\bar{x})=-\frac{\delta_i^2}{2}\sin^2(\kappa\bar{x})+\frac{3}{8}\delta_i^3\sin(\kappa\bar{x})\sin(2\kappa\bar{x}).
\label{eq15}
\enea

  \end{section}

  \begin{section}{Conclusion}
%

In conclusion, we have obtained some new and interesting results in discussing the problem
of nonlinear electron plasma oscillation by considering the relativistic effect on 
inhomogeneous plasma system. A modified expression for the frequency of the characteristic plasma wave indicating 
a shifting has been derived in order to study the phase mixing process leading to wave breaking. It is observed from the expression 
of the phase mixing time that combined effect of inhomogeneity
and relativity expedite the process of phase mixing. Such theoretical investigation could have some practical implications 
in the plasma based acceleration process where the electric field of the plasma wave 
interacts with the highly relativistic particles. To the best of our knowledge, we are 
reporting here for the first time the effect of background density inhomogeneity on the relativistic plasma wave.

%
%

%
%

\end{section}


\begin{thebibliography}{99}



\bibitem{akhiezer} A. I. Akhiezer and R. V. Polovin, Zh. Eksp. Teor. Fiz. {\bf 30}, 915 (1956) [Sov.
Phys. JETP {\bf 3}, 696 (1956)].
\bibitem{davidson1} R. Davidson and P. Schram, Nuclear Fusion {\bf 8} 183 (1968).
\bibitem{davidson2} R. C. Davidson, Methods in Nonlinear Plasma Theory (Academic, New York, 1972).
\bibitem{davidsonbook} R. C. Davidson, \textit{Methods in Nonlinear Plasma Theory} (Academic, New York, 1972).
\bibitem{stenflo} L. Stenflo, and G. Brodin, Phys. Plasmas {\bf 23}, 074501 (2016).
\bibitem{brodin} G. Brodin, and L. Stenflo, Phys. Plasmas {\bf 24}, 124505 (2017).
\bibitem{xu} H. Xu, Z. M. Sheng, and J. Zhang, Phys. Scr {\bf 74}, 673 (2006).
\bibitem{zhou} S. Zhou, Phys. Scr {\bf 91}, 025601 (2016).
\bibitem{joshi} C. Joshi, W. B. Mori, T. Katsouleas, J. M. Dawson, J. M.
Kindel, and D. W. Forslund, Nature (London) {\bf 311}, (1984) 525.
\bibitem{modena} A. Modena, Z. Najmudin, A. E. Dangor, C. E. Clayton, K. A. Marsh, C. Joshi,
V. Malka, C. B. Darrow, C. Danson, D. Neely, and F. N. Walsh, Nature (London) {\bf 377}, (1995) 606.
\bibitem{tajima} T. Tajima and J. M. Dawson, Phys. Rev. Lett. {\bf 43}, 267 (1979).
\bibitem{dawson} J. M. Dawson, Phys. Rev. {\bf 113}, 383 (1959).
\bibitem{coffey} T. P. Coffey, Phys. Fluids {\bf 14}, 1402 (1971).
\bibitem{katsouleas} T. Katsouleas and W. B. Mori, Phys. Rev. Lett. {\bf 61}, 90 (1988).
\bibitem{sengupta2} S. Sengupta, V. Saxena, P. K. Kaw, A. Sen, and A. Das, Phys. Rev. E {\bf 79}, 026404 (2009).
\bibitem{sengupta3} S. Sengupta, P. K. Kaw, V. Saxena, A. Sen, and A. Das,  Plasma Phys. Controlled Fusion {\bf 53}, 074014 (2011).
\bibitem{pramanik} S. Pramanik, C. Maity, and N. Chakrabarti, Phys. Plasmas {\bf 22}, 052303 (2015).
\bibitem{pramanik2} S. Pramanik, C. Maity, and N. Chakrabarti, Phys. Scr {\bf 91}, 065602 (2016).
\bibitem{infeld1} E. Infeld, and G. Rowlands, Phys. Rev. Lett. {\bf 62}, 1122 (1989).
\bibitem{nappi} C. Nappi, A. Forlani, and R. J. Fedele, Phys. scr. {\bf 43}, 301 (1991).
\bibitem{infeld2} E. Infeld, G. Rowlands, and S. Torven , Phys. Rev. Lett. {\bf 62}, 2269 (1989).
\bibitem{infeld3} E. Infeld, and G. Rowlands, Physical Review A {\bf 14}, 838 (1990).
\bibitem{kaw} P. K. Kaw, A. T. Lin, and J. M. Dawson, Phys. Fluids {\bf 16}, 1967--1975 (1973).
\bibitem{bogoliubov} N. N. Bogoliubov and V. A. Mitropolsky, Asymptotic Methods
in the Theory of Nonlinear Oscillations, Hindustan Publishing, Delhi (1961).

 
%
%
%
%



%
%
%
%
%
%
%
%
%
%
%
%





 



\end{thebibliography}
\end{document}